# A comparative study of the magnetic and magnetotransport properties between a metallic (x=0.6) and a semiconducting (x=0.2) member of the solid solution LaNi$_x$Co$_{1-x}$O$_3$.

J. Androulakis[a,b], N. Katsarakis[a], Z. Viskadourakis[a], J. Giapintzakis[a,c]

[a]Institute of Electronic Structure and Laser, Foundation for Research and Technology –
Hellas, P.O. Box 1527, Vasilika Vouton, 711 10 Heraklion, Crete, Greece

[b]Department of Chemistry, University of Crete, Leoforos Knossou, 714 09 Heraklion, Crete,
Greece

[c]Department of Materials Science and Technology, University of Crete,
P.O. Box 2208, 710 03, Heraklion, Crete, Greece

## Abstract

We present a comparative study of both the magnetic and magnetotransport properties for two members of the perovskite solid solution LaNi$_x$Co$_{1-x}$O$_3$ (x=0.2, 0.6) located on opposite sides of the chemically induced metal-to-insulator transition. LaNi$_{0.6}$Co$_{0.4}$O$_3$ exhibits metallic behavior and small but negative magnetoresistance, whereas LaNi$_{0.2}$Co$_{0.8}$O$_3$ exhibits semiconducting behavior and giant negative magnetoresistance at low temperatures. On the other hand, we observe pronounced similarities in the magnetic properties of both compounds. A consistent explanation regarding the origin of the magnetoresistance in the two members of the solid solution is provided.

# I. Introduction

Transition metal oxides with the perovskite structure have attracted a lot of attention the past few years due to their unusual magnetic and transport properties, which include high temperature superconductivity in the cuprates and colossal magnetoresistance in the manganites. The studies of cuprates and manganites have revealed a plethora of interesting physical phenomena and have spurred up interest regarding potential technological applications [1,2].

The members of the $LaNi_xCo_{1-x}O_3$ solid solution are structurally related to the above mentioned systems and also exhibit a variety of interesting physical phenomena such as metallic conductivity for $x>0.35$ and giant negative magnetoresistance (GMR) for $x<0.35$ [3]. The remarkable phenomenon of metallic conductivity in this series has been the subject of several studies and has already been considered for practical applications, e.g. cathode materials in oxide fuel cells [4]. On the other hand, the appearance of GMR seems to lack a consistent explanation. Perez et al. have attributed the existence of GMR in the $x=0.3$ member of the series to a spin glass phase [5]. However, recent detailed ac and dc magnetic susceptibility measurements seem not to support the proposed picture [6].

The end members of the series are: $LaNiO_3$, a Stoner enhanced metallic paramagnet [7], and $LaCoO_3$, an insulator that exhibits remarkably complex magnetic behavior [8]. The substitution of Co by Ni introduces disorder as the end members of the series have slightly different lattice parameters. The usual description of the electronic structure for this system is based on the hybridisation of oxygen p orbitals with $e_g$ transition metal orbitals to form the $\sigma^*$ conduction band; a rather simplistic framework as it neglects electron-electron correlations to a large extend [9].

Early investigations of the magnetic properties of the metallic members of the solid solution have reported the existence of ferromagnetism with a glassy nature [10], but have not fully explored its origin. However, strong ferromagnetic (FM) signals, and metastable magnetic phenomena are also known to occur for the semiconducting compositions [5, 6, 11].

In view of the above, the purpose of the present work is to make a comparative study of both the magnetic and magnetotransport properties for two members of the series, one of which belongs to the metallic regime (x=0.6) and the other to the semiconducting regime (x=0.2). Both members are away from the critical composition, $x_c$=0.35, at which the metal-to-insulator transition occurs, in order to avoid effects related to such a critical ground state. We observe pronounced similarities in the magnetic properties and distinct differences in the magnetoresistance (MR) of both compounds. A consistent picture regarding the origin of the MR in the two members of the $LaNi_xCo_{1-x}O_3$ solid solution is provided.

## II. Experimental Part

$LaNi_xCo_{1-x}O_3$ powder samples were prepared by a coprecipitation method, using high purity (99.999 %) metal nitrates ($La(NO_3)_3 \cdot 6H_2O$, $Co(NO_3)_2 \cdot 6H_2O$, $Ni(NO_3)_2 \cdot 6H_2O$) as starting materials. The nitrates were first dissolved in water. Coprecipitation of mixed lanthanum, cobalt and nickel hydroxides was achieved by adding aqueous NaOH solution as a precipitating agent keeping pH~11. The precipitates were carefully washed, dried and finally decomposed at 1073 K. The precursor powders obtained were subsequently heated in air at 1273 K for one day and then slowly cooled down to room temperature.

Powder X-Ray diffraction measurements were performed at room temperature using a Rigaku powder X-Ray diffractometer (RINT-2000). The samples were found to be free of impurity phases within the resolution of the instrument. Profile analysis performed in the hexagonal system yielded the correct space group and lattice parameters that were in agreement with previously published work [5, 12]. Energy Dispersive X-ray (EDX) analysis confirmed the correct cation stoichiometry and iodometric titrations showed that the produced powders were optimally oxygenated. Magnetic measurements were performed on powder samples in the temperature range $2 < T < 280$ K using a MagLab EXA Susceptometer by Oxford Instruments. Transport data were obtained from bar-shaped pellets with the magnetic field applied perpendicular to the current direction using a home-made apparatus. Data for the metallic (x=0.6) sample were collected using a four-probe low frequency (20 Hz) technique with the use of a lock-in amplifier in order to eliminate thermal emf effects whereas the data for the semiconducting sample were collected using a two-probe bipolar technique employing an electrometer.

## III. Results

*dc Susceptibility*

Fig. 1 shows the low field static susceptibility, $\chi_{dc}$, for both compounds. The data were collected under zero field cooled (ZFC) and field cooled (FC) conditions, upon warming the samples in both cases at a constant rate of 0.3 K/min. Both compounds exhibit thermomagnetic irreversibility with qualitatively similar behavior. Specifically, there is a sudden increase in $\chi_{dc}$ below 60 K for both compounds, followed by a peak at $T_p \sim 32$ K for x=0.2 and at $T_p \sim 28$ K for x=0.6, a minimum at lower temperatures and a subsequent increase below 10 K for all curves. In addition, the irreversibility temperature, $T_{irr}$, is greater than $T_p$ for both compounds ($T_{irr}\sim38$ K for x=0.2; $T_{irr}\sim50.5$ K for x=0.6).

The shoulder designated by the arrow around 55 K in the susceptibility curve of x=0.6 is believed to be an effect of the Ni sublattice. According to Vergard's law for solid solutions, it is reasonable to expect the Ni sublattice to have an increasing impact on the magnetic properties with increasing Ni concentration. Moreover, the susceptibility data in the temperature range 30-55 K are fitted relatively well by the formula, $\chi(T)=\chi(0)-\alpha T^2+C/T$ , which has been previously used to fit susceptibility data for $LaNiO_3$ [7]. Finally, it is worth mentioning that in Fig. 4 and 5 of ref. 7 it is clearly shown that the susceptibility of $LaNiO_3$ exhibits an enhancement around 50 K and another anomaly at 12 K.

The effect of increasing magnetic field on the static susceptibility of $x=0.6$ is shown in Fig. 2. Thermomagnetic irreversibility effects are steadily suppressed while the upturn at low temperatures is gradually vanishing. Similar effects are also observed for the x=0.2 semiconducting compound [6]. Notice that the peak at 55 K, which was observed on the low field static susceptibility curves of x=0.6, does not appear at higher fields.

The large increase of $\chi_{dc}$ around 60 K for both compounds is an indication for the existence of FM correlations, which were certified by the measurement of hysteresis loops (vide infra). Fits to the Curie-Weiss law, $\chi_{dc}^{-1}=(T-\Theta_C)/C$, for $80 < T < 280$ K gave positive mean field $\Theta_C$ values of $55.6 \pm 1.5$ K for x=0.2 and $51.1 \pm 1.3$ K for x=0.6. The positive $\Theta_C$ values are consistent with the existence of FM interactions. The spin-only value of the effective magnetic moment obtained from the above fit is $\mu_{eff}=2.39 \pm 0.06$ $\mu_B$ for x=0.2 and $\mu_{eff}=2.20 \pm 0.06$ $\mu_B$ for x=0.6.

The divergence of the magnetic susceptibility at the Curie temperature, $T_C$, has been studied using the Kouvel-Fisher (KF) scaling analysis [13[, which makes use of the scaling law:

$$\chi \propto (T-T_C)^{\gamma} \qquad (1)$$

and defines a new function:

$$T_{KF} = \frac{1}{\chi(d\chi^{-1}/dT)} \qquad (2)$$

to determine $T_C$ and $\gamma$:

$$T_{KF} = \frac{T-T_C}{\gamma} \qquad (3)$$

The KF scaling analysis for x=0.6 is presented in Fig. 3. The parameters obtained are $T_C$=55.7±5.68 K and $\gamma$=1.05±0.06. The same analysis for x=0.2 yielded $T_C$=55.31±5.64 K and $\gamma$=0.54±0.06. The $T_C$ values obtained from the above analyses are in agreement with the calculated mean field $\Theta_C$ values. Just for completeness, we mention that the value of the critical exponent $\gamma$, which describes the divergence of the susceptibility just above $T_C$, for the mean field approach to the Landau theory is $\gamma$=1, for a 3D Isisng model is $\gamma$=1.25 and for a spin-1/2 Heisenberg model is $\gamma$=1.43 [14].

Fig. 4 presents M(H) curves for both compounds measured at T=2 K. The observation of hysteretic behavior firmly establishes the existence of FM interactions. The magnetization does not reach saturation up to the maximum applied field of 30 kOe, which we atribute to the spin -disorder present in the measured compounds. The calculated magnetization at H=30 kOe and T=2 K is 0.44 $\mu_B$/f.u. for x=0.2 and 0.26 $\mu_B$/f.u. for x=0.6. A distinctive difference between the M(H) curves is the relatively large coercive field of the semiconducting  compound in comparison to the metallic one. The coercive field, $H_c$, for x=0.6 reaches $H_c \sim 200$ Oe, while for x=0.2, $H_c \sim 1.6$ kOe.

*ac Susceptibility*

Fig. 5 shows the temperature dependence of the ac susceptibility, $\chi_{ac}(T)$, for different values of the applied dc magnetic field. Measurements were made upon warming the ZFC samples at a constant rate of 0.3 K/min and using an ac driving field of 1 Oe oscillating at 1 kHz. In zero dc field both compounds exhibit a broad featureless peak with a maximum in the range 35-40 K (Fig. 5a). In the presence of dc field the single peak evolves into two separate peaks. The magnitude of the two peaks is reduced drastically with increasing dc field intensity (see Fig. 5b and 5c). Note also, that the zero field peak at 55 K, for x=0.6, disappears in moderate dc fields consistent with the dc data.

Fig. 6 exhibits the frequency dependence of $\chi_{ac}(T)$ for the x=0.6 compound in zero dc field. The overall behavior of $\chi_{ac}(T)$ with respect to frequency around 35 K is consistent with that of glassy systems and similar to the one that has been reported for x=0.2 and 0.3 [5, 6, 11]. Nevertheless, thermomagnetic irreversibility of $\chi_{dc}$ and frequency dependence of $\chi_{ac}(T)$ are not sufficient evidence to safely conclude on the origin of glassy dynamics, since similar effects have been observed even in systems which exhibit long range magnetic ordering [15]. A detailed study of the non-linear components of $\chi_{ac}(T)$ could shed more light on this issue.

Both the second, $\chi_2$, and the third, $\chi_3$, harmonic components of $\chi_{ac}(T)$ for x=0.2 and 0.6 compounds have been measured. (Fig. 7 depicts the temperature dependence of $|\chi_2 h_0|$ and $|(3/4)\chi_3 h_0^2|$ for x=0.2). Both compounds exhibit a strong signal of $\chi_2$, which pertains to the presence of a spontaneous moment [6, 16]. In addition, $\chi_3$ for both compounds does not exhibit any divergence and thus cannot be attributed to a random freezing of atomic local moments, i.e., a canonical spin glass state. We should point out that similar non-diverging behavior of $\chi_3$ has also been observed in superparamagnetic (SP) systems [17]. To check for the presence of SP in x=0.2, we have measured M(H) curves in the range $30 < T < 50$ K (not shown here) where SP behavior is expected to appear. Experimental M(H) isotherms do not scale with H/T, thus excluding the presence of SP and pointing rather to the existence of strong FM correlations. In addition, we have neither observed a $T^{-3}$ dependence of $\chi_3$ above $T_f$ (defined as the peak in the zero field ac susceptibility) nor a temperature independence of both $\chi_1$ and $\chi_3$ below $T_f$, which are expected by Wohlfarth's model for SP [17]. A similar analysis for the x=0.6 compound led to the same conclusion, i.e., it does not exhibit SP behavior but a rather disordered spontaneously magnetizing state.

Evidently, the overall magnetic behavior is strikingly similar for both compounds, which suggests that possibly the magnetic interactions involved on both sides of the chemically induced metal-to-insulator transition are of the same origin. Our data seem to exclude the presence of both a conventional spin glass phase as well as the formation of SP clusters in the vicinity of 55 K and instead, suggest the formation of magnetic clusters with enhanced FM intra-cluster correlations. Furthermore, it is remarkable that the substitution of an extra 40% of Co by Ni modifies only slightly almost all of the magnetic quantities examined so far, i.e., the $T_C$ value (obtained by KF analysis), the mean field $\Theta_C$ value, the spin-only $\mu_{eff}$ value as well as the low-temperature magnetization [M(30 kOe) at 2 K]. An exception to the above quantities is the magnitude of the coercive field, which drops an order of magnitude from x=0.2 to x=0.6. However, when we also take into consideration the available transport data, this $H_c$ effect has a natural explanation, which will be discussed in section IV.

*Resistivity*

Fig. 8 shows the temperature dependence of the zero-field resistivity for the x=0.2 compound, which exhibits a semiconducting behavior. In general, the semiconducting transport data are modelled using one of the three theoretical frameworks: (a) The Arrhenius law, $\rho=\rho_0 \exp\{-E_g/k_B T\}$, which is used to describe thermally activated behavior due to a band gap $E_g$, (b) nearest neighbor hopping of small polarons, which is described by the formula $\rho=\rho_0 T\exp\{-E_p/k_B T\}$, with $E_p$ being the characteristic energy for polaron hopping and (c) Mott's variable range hopping (VRH) model described by the expression $\rho=\rho_0 \exp\{-T_0/T\}^{1/4}$. $T_0$ is related to the characteristic fall-off rate of the radial part of the electronic wavefunction around an impurity ion, $\alpha$, through the relation: $k_B T_0=1.5/\alpha^3 N(E_F)$, where $N(E_F)$ is the density of states at the Fermi level [18]. The VRH model originates from the effect of high doping that strongly disorders a semiconducting crystal giving rise to the Anderson localization phenomenon [19].

The inset of Fig. 8 shows a $\ln\rho$-$T^{-1/4}$ plot of the data; the curve is linear from 5 to 70 K. It is interesting to mention that $\ln(\rho/T)$-$(1/T)$ and a $\ln\rho$-$(1/T)$ plots exhibit no linearity throughout the whole temperature interval of the measurements (not shown here). Therefore, we can safely conclude that the transport properties of the semiconducting phase, at least down to 5 K, are strongly dominated by the effect of Anderson localization, which introduces a mobility edge above $E_F$ that stabilizes the semiconducting phase. An earlier study of the $x=x_c$ member of the $LaNi_xCo_{1-x}O_3$ solid solution mentioned the existence of a limited temperature range for which the VRH model is valid, however, the authors did not extend their analysis to x<$x_c$ [20]. It should be mentioned that transport studies performed on x=0.1 and x=0.5 single crystals concluded that the data are typical of small polaron hopping [21]. However, their conclusion seems to be based on qualitative comparison of their data with those of the LaSr$_x$Co$_{1-x}$O$_3$ system and not from a quantitative fitting of the actual transport data to the small polaron hopping model.

The temperature dependence of the resistivity of the metallic sample (x=0.6) is depicted in Fig. 9. At high temperatures we observe a linear increase of $\rho(T)$ which is consistent with the metallic nature of x=0.6. On the other hand, the low temperature behavior is characterized by an upturn, which has been interpreted as the effect of strong electron-electron correlations in the oxide [9]. Apparently, there is no simple formula that fits the $\rho(T)$ curve. In the past, the very low temperature resistivity of the members of the solid solution with x>0.4 has been fitted relatively well by a $\sim T^{1/2}$ term which quantifies electron-electron correlations plus a power law $\sim T^p$ term to account for inelastic contributions at higher temperatures [22]. The solid line of Fig. 9 is the result of a fit to the resistivity data using the following expression:

$$\rho(T) = \rho_0 + \alpha_1 T + \alpha_2 T^{1/2} - \alpha_3 \ln T \qquad (4)$$

The expression used contains three terms besides the residual resistivity term, $\rho_0$ (where $\rho_0 = \rho(T=0)$): (i) a linear term effective at high temperatures to account for phonon scattering, (ii) a $\sim T^{1/2}$ term that quantifies strong electron-electron interactions at low temperatures and (iii) a $\sim \ln T$ term which normally arises in systems dominated by spin fluctuations [23]. We believe that the lnT term plays a decisive role in determining the behavior of the magnetoresistance of the metallic phase. The inset of Fig. 9 depicts low temperature (T<50 K) resistivity data taken under 0 and 20 kOe as a function of lnT. The clear suppression of the observed linear part (decreasing slope with increasing field) is a strong indication that the main contribution to the magnetoresistance is the suppression of spin fluctuations. Preliminary data on x=0.7, i.e., deeper in the metallic state, provide further support to the above argument.

The effect of the applied magnetic field on the resistivity of both compounds is depicted in Fig. 10. It is readily seen that both of them exhibit negative MR at 5 K. However, MR for x=0.2 reaches 60% at the maximum field applied (65 kOe) whereas the maximum MR value for x=0.6 does not exceed 6%. Notice also that for x=0.2 the MR values at low fields are positive; in the absence of any suitable theoretical model we do not further discuss this effect. However, such a behavior certainly deserves further consideration and it would be useful to examine the magnetotransport properties for $x<x_c$ by means of Hall effect as well as thermopower measurements under applied magnetic field.

## IV. Discussion

As it has already been mentioned in the introduction, the GMR phenomenon has emerged as a rich and extremely active topic of experimental research. The underlying mechanisms, which are responsible for the appearance of GMR in different systems may be of completely different origin. In the manganites, for example, the subtle interplay of charge, spin and orbital degrees of freedom along with Zener's double exchange mechanism are widely considered responsible for the emerging GMR effect at and around the FM Curie temperature [2]. On the other hand, a large number of transition metal oxides, which exhibit large or giant negative MR, largely remain beyond our understanding. Regarding the investigated system, it has been previously suggested that it exhibits similar behavior to the magnetoresistive manganites [5]. However, our study reveals that there are distinct differences between the investigated system and the manganites. For example the x=0.2 compound exhibits GMR which is enhanced with decreasing temperature without a temperature induced metal-to-insulator transition. Such effects are certainly worth of further consideration.

Next we proceed to discuss our experimental results beginning with the x=0.2 magnetoresistive compound. As it has been previously mentioned, the dominant transport mechanism in the temperature range of interest is VRH. The random potential fluctuations needed for VRH are introduced by replacing Co with Ni. The introduction of Ni into the lattice strongly disorders the $LaCoO_3$ matrix as it introduces one extra $e_g$ electron. In addition, it has been shown that Co ions, in $LaCoO_3$, can easily obtain higher spin states (intermediate spin (IS): $t_{2g}^5 e_g^1$ and high spin (HS): $t_{2g}^4 e_g^2$) in the presence of magnetic impurities [24]. Whatever the spin state of Co, we expect Ni-doping to introduce carriers, i.e., holes in the $\pi$ ($t_{2g}$) band on nearest neighbor Co sites, a suggestion consistent with thermopower data [25]. We believe that the energy difference between Ni and Co 3d-orbitals, which is of the order of ~1 eV [9], plays a decisive role in localizing the Co-related carriers around Ni, thus forming bound magnetic polaron-like clusters. This mechanism would be enhanced both by charge disproportionation of Ni [5], which presumes two $e_g$ electrons per Ni site, as well by the coalescence of Ni ions.

The characteristic temperature $T_0$ was calculated to be of the order of ~$10^5$ K, a value physically unjustifiable. However, we should mention that large $T_0$ values have also been previously reported for lanthanum cobaltate perovskite oxides [26, 27]. We believe that unusually high $T_0$ values are related to the small bandwidth exhibited by this type of oxide-semiconductors. Furthermore, we should point out that at sufficiently low temperatures, T<20 K, a modified version of Mott's VRH can fit the data equally well (not shown here). Efros and Shklovskii showed that the Coulomb repulsion between carriers leads to a form of hopping for which $\rho=\rho_0\exp\{-T_0/T\}^{1/2}$ [28]. In this case, $T_0'$ is related to the average localization length, $\xi$, through the relation $k_BT_0=1.5e^2/\kappa\xi$, where $\kappa$ is the dielectric constant. Using the above relation, we can roughly estimate $\xi$ to be ~3 A. (For this estimation we used $\kappa$~10, which is a typical value for semiconductors). It is worth mentioning that Gayathri et al. concluded that Co doping of $La_{0.7}Ca_{0.3}MnO_3$ destroys metallicity along with magnetic long range ordering and leads to the formation of FM clusters, while the carriers exhibit an average localization lenght of 2.5-5 A [27]. Therefore, the divergence of the magnetic susceptibility at $T_C$ could be attributed to the confinement of induced carriers around magnetic Ni ions upon lowering the temperature below $T_C$, and that these carriers order ferromagnetically in order to minimize their energy.

Next we present a plausible scenario within which we attempt to explain the magnetic as well as the transport data for both compounds. We propose that there are randomly dispersed Ni-based clusters and the intra-cluster exchange interactions between the Ni local moments and the induced carriers are FM. Furthermore, we attribute the observed metastable phenomena to the random distribution of the cluster moments in all directions. In the low-Ni-doping regime, the magnitude of the localization length (~3 A) indicates that the formed FM clusters most likely exhibit single domain-like behavior. In general, the magnetization of such a small cluster in the presence of an applied field should rotate as a whole, a process that usually requires large fields. Thus, we believe that the magnitude of the coercive field for x=0.2 reflects the small size of the cluster moments in the semiconducting matrix. On the other hand, the smaller coercive field for the metallic sample indicates that the size of the magnetic clusters is larger compared to that of the clusters in the semiconducting sample, and hence the rotation of the magnetization can be explained based on boundary displacement. The metallic nature of the resistivity suggests that these magnetic clusters should be treated as enhanced spin concentration micro-regions which act as scattering centers in the $\sigma^*$ delocalized conduction band formed by the extended overlap of Ni antibonding $e_g$ orbitals. The appearance of the lnT term in eq. (4) is consistent with the proposed spin fluctuation scattering mechanism.

The large negative values of MR for x=0.2 are typical of samples exhibiting hopping conduction involving cluster-like bound magnetic polarons [29]. The increasing magnetic field results in magnetizing the $Co^{3+}$ matrix, i.e., Co obtains IS or HS configurations, because, along with the magnetic ions, it strongly perturbs the relation $\Delta_{cf} \approx \Delta_{ex}$, which holds for $LaCoO_3$ [8, 30] ($\Delta_{cf}$ is the crystal field splitting energy and $\Delta_{ex}$ is the intra-atomic exchange energy for $Co^{3+}$). The reduction of the difference in the magnetization between the cluster polaron and the matrix leads to a spatial expansion of the orbits of the trapped charge carriers, i.e., the spatial hopping range increases, and this leads to GMR. On the other hand, the MR values are not expected to drop significantly in the x=0.6 metallic sample, because the percolation conduction of metallic clusters leaves unaffected the mean free path upon applying a magnetic field. Thus, small but still negative MR values most likely result by the suppression of spin fluctuations below the characteristic FM temperature (see inset of Fig. 9).

### V. Conclusions

We have undertaken a comparative study between a semiconducting (x=0.2) and a metallic (x=0.6) member of the solid solution $LaNi_xCo_{1-x}O_3$. $LaNi_{0.6}Co_{0.4}O_3$ is a strongly correlated metal and exhibits small but negative MR whereas $LaNi_{0.2}Co_{0.8}O_3$ exhibits semiconducting behavior and GMR at low temperatures. Both compounds exhibit strikingly similar magnetic properties, which are attributed to the formation of a cluster-glass-like magnetic phase with strong FM intra-cluster interactions. Based on magnetotransport data, the negative but small MR of the metallic sample, x=0.6, is suggested to arise from the suppression of spin fluctuations while the GMR observed for x=0.2 is attributed to the formation of magnetic clusters which grow spatially inside the matrix with increasing applied magnetic field.

**Figure Captions**

**Fig.1** Temperature dependence of the molar dc susceptibility for both compounds in low field, $H_{dc}$=100 G. ($\diamond$) ZFC data for x=0.2 ($\triangle$) FC data for x=0.2 ($\bullet$) ZFC data for x=0.6 ($\circ$) FC data for x=0.2. The shoulder indicated by the arrow is most likely an effect of the Ni sublattice.

**Fig.2** Temperature dependence of the dc molar susceptibility for $LaNi_{0.6}Co_{0.4}O_3$ measured under both ZFC nad FC conditions in several magnetic fields ($0.8 < H < 10$ kOe). Notice that the thermomagnetic irreversibility is gradually vanishing with increasing field.

**Fig.3** KF function (in arbitrary units) versus temperature for the metallic sample (x=0.6). The high-temperature fit defines both $T_C$ and the critical exponent, $\gamma$.

**Fig.4** Magnetic field dependence of the magnetization for x=0.2 (a) and x=0.6 (b) at 2 K. Note the enhanced value of the coercive field for x=0.2.

**Fig.5** Field dependence of the real part of the linear ac magnetic susceptibility, $\chi_{ac}$ , as a function of temperature for x=0.2 (solid line) and x=0.6 (dashed line). Data were taken with a driving ac field of amplitude $h_0$=1 Oe oscillating at a frequency of 1 kHz [(a) $H_{dc}$=0 kOe, (b) $H_{dc}$=1 kOe, (c) $H_{dc}$=5 kOe]

**Fig.6** Temperature dependence of the real part of the linear ac magnetic susceptibility, $\chi_{ac}$ , for x=0.6 at several frequencies. Data were taken in zero dc magnetic field with an ac field of 1 Oe oscillating at 1 kHz.

**Fig.7** Temperature dependence of the absolute value of the second, $|\chi_2 h_0|$, and the third, $\frac{3}{4}|\chi_3 h_0^2|$, harmonic components for x=0.2 compound extracted from the signal obtained at frequency $2\omega$ and $3\omega$, respectively ($\omega=2\pi$ f, f=1 kHz).

**Fig. 8** Resistivity data for the x=0.2 compound in the temperature range 5< T < 70 K in zero applied field. Inset: the solid line is the fit to the resistivity data based on Mott's VRH model.

**Fig. 9** Resistivity data for the x=0.6 compound (open circles) in the temperature range 5 < T < 300 K in zero applied field. The solid line is the fit to the data using equation (4). The maximum calculated fitting error using the relation 100 $[(\rho_{obs}-\rho_{fit})/\rho_{obs}]$ ($\rho_{obs}$=observed resistivity, $\rho_{fit}$=calculated resistivity using equation (1)) is 0.6%. Inset: $\rho$ vs lnT plots and linear fittings for $H_{dc}$=0 and 20 kOe. Notice the suppresion of the lnT term (decreasing slope of the linear part) with increasing magnetic field.

**Fig. 10** (a) Magnetoresistance data, defined as 100 $[(\rho(H)-\rho(0))/\rho(0)]$, for the indicated compounds at 5 K. All data were collected with the applied field perpendicular to the current direction. (b) The temperature dependence of the MR=100 $[(\rho(65kOe)-\rho(0))/\rho(0)]$ for the x=0.2 compound.

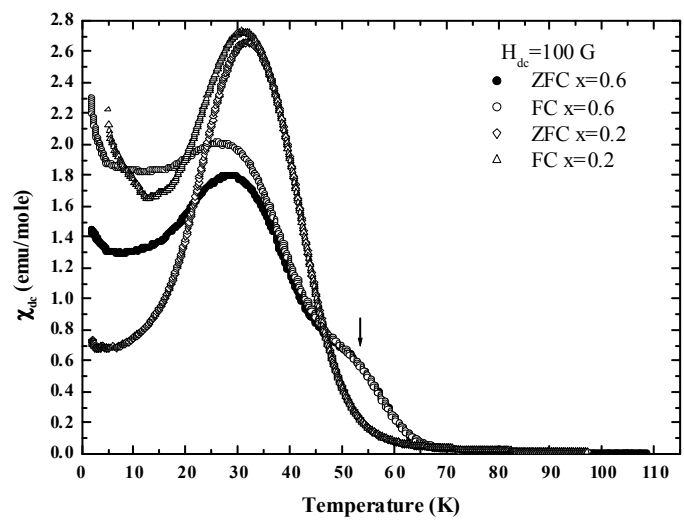

**Figure 1**

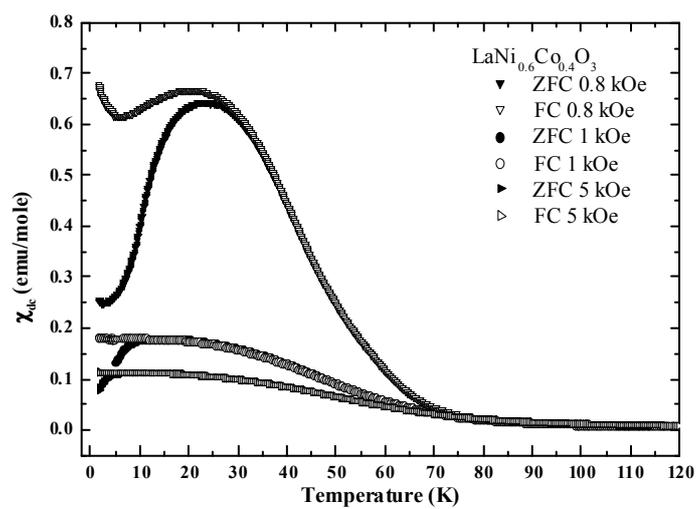

**Figure 2**

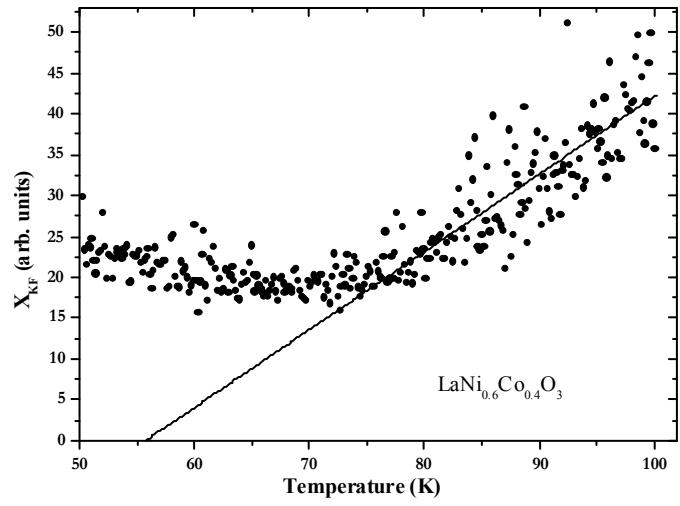

**Figure 3**

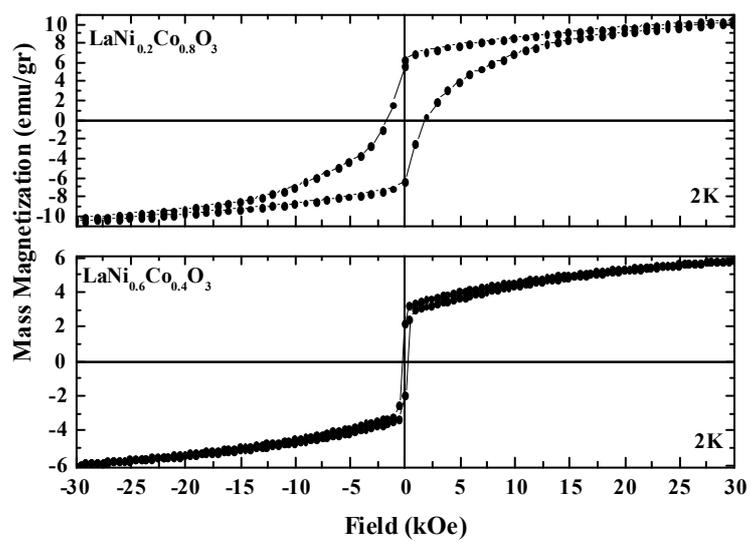

Figure 4

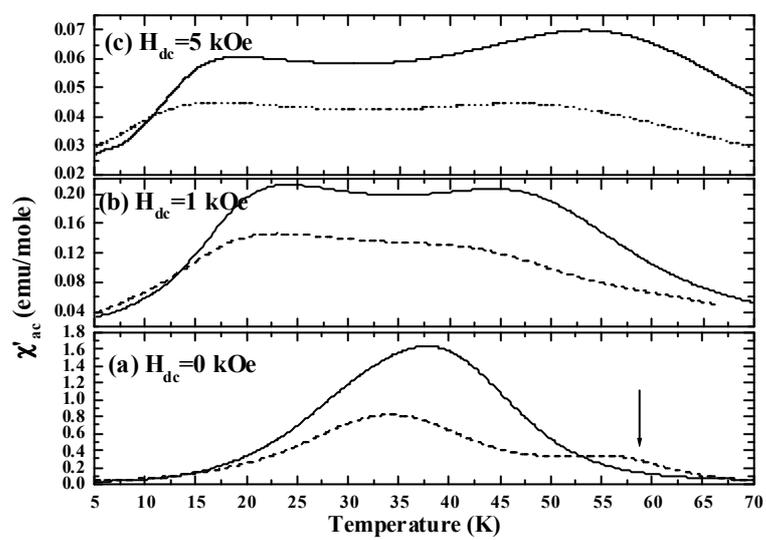

**Figure 5**

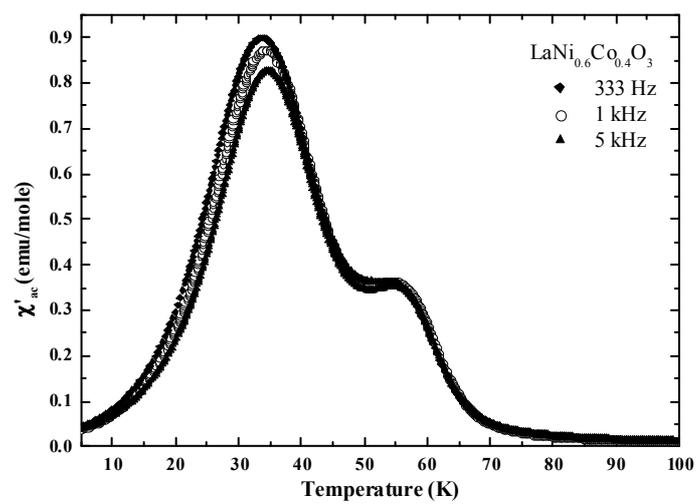

**Figure 6**

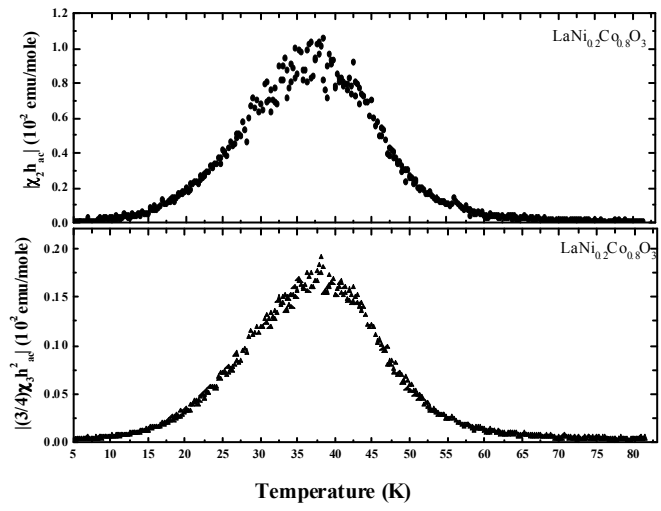

**Figure 7**

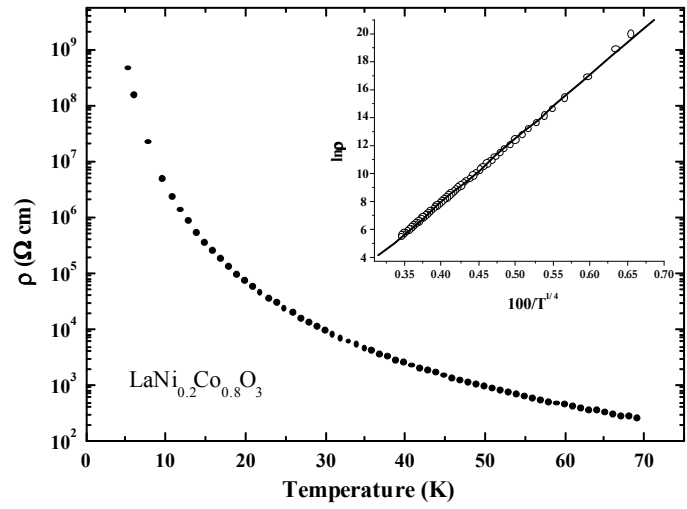

**Figure 8**

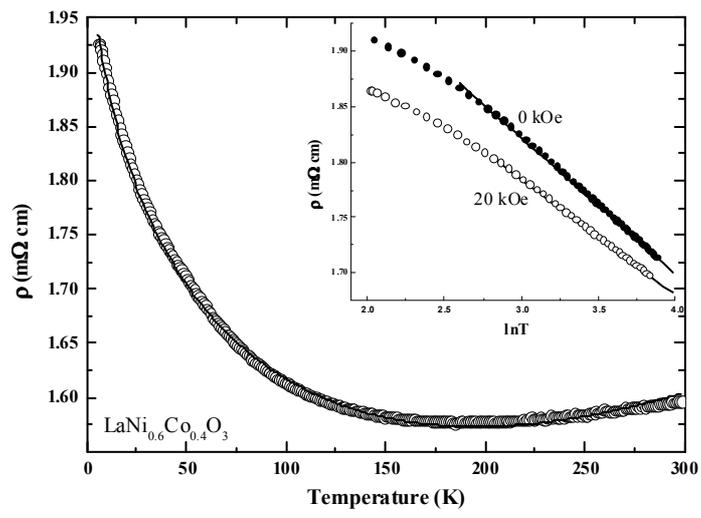

**Figure 9**

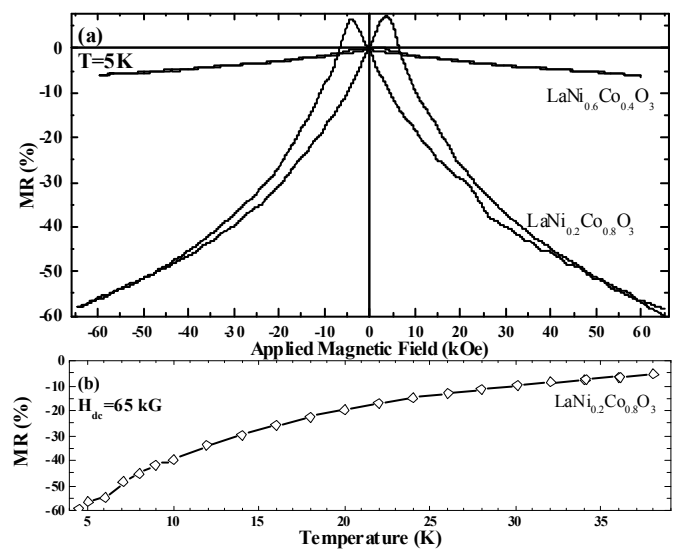

**Figure 10**